\documentclass[aps,prl,twocolumn,amsmath,amssymb]{revtex4}
\usepackage{graphicx}
\usepackage{bm}

\begin{document}

\title{
Solid-State Quantum Communication With Josephson Arrays}
\author{Alessandro Romito, Rosario Fazio} 
\affiliation{NEST-INFM $\&$ Scuola Normale Superiore, Piazza dei Cavalieri 7, 
56126 Pisa, Italy}
\author{C. Bruder}
\affiliation{Department of Physics and Astronomy,
University of Basel, Klingelbergstrasse 82, 4056 Basel, Switzerland}

\begin{abstract}
Josephson junction arrays can be used as quantum channels to transfer
quantum information between distant sites. In this work we discuss
simple protocols to realize state transfer with high fidelity.  The
channels do not require complicate gating but use the natural dynamics
of a properly designed array. We investigate the influence of static
disorder both in the Josephson energies and in the coupling to the
background gate charges, as well as the effect of dynamical noise. We
also analyze the readout process, and its backaction on the state transfer.
\end{abstract}
\date{\today}
\maketitle

The transmission of a quantum state through a channel between distant
parties is an important issue in quantum communication.  In optical
systems photons can be transferred coherently over large
distances~\cite{gisin}.  However, it is also highly desirable to have
similar protocols for quantum information transfer in solid-state
environments. A possible solution would be to interface solid-state
quantum hardware to optical systems~\cite{zoller}. Another possibility
is to use flying qubits, i.e. to transfer the physical qubits
along leads~\cite{burkard2000}.  
Inspired by the paper of Bose~\cite{bose03} the idea of our  work is to 
construct a genuine quantum transmission line using a Josephson junction 
array.

Recently, a spin chain with ferromagnetic Heisenberg interactions has
been proposed for quantum communication~\cite{bose03}.  It was shown that
Heisenberg chains can be used to transfer unknown quantum states over
appreciable distances ($\sim 10^2$ lattice sites) with high
fidelity~\cite{bose03,subrahmanyam03,giovannetti04,plenio04}. 
By preparing the state 
to be transferred at one end of the chain and waiting for a well-defined
time interval, one can reconstruct the state at the other end of the
chain. Even perfect transfer could be achieved over arbitrary
distances in spin chains~\cite{perfecttransfer}. 
Quantum state transport through harmonic chains was considered in
Ref.~\onlinecite{eisert}.

Josephson qubits are among the most promising candidates as building
blocks of quantum information
processors~\cite{schoenreview,averinreview}.  In this Letter, we
extend their application range to quantum communication and show that
a one-dimensional Josephson array is a natural transmission line for
systems with superconducting charge qubits.  We calculate the
transmission fidelity and investigate the effect of
static inhomogeneities and dynamical noise. We also analyze the
readout process by a single-electron transistor (SET) at the end of the array.

\begin{figure}[thb]
\includegraphics[width=70mm]{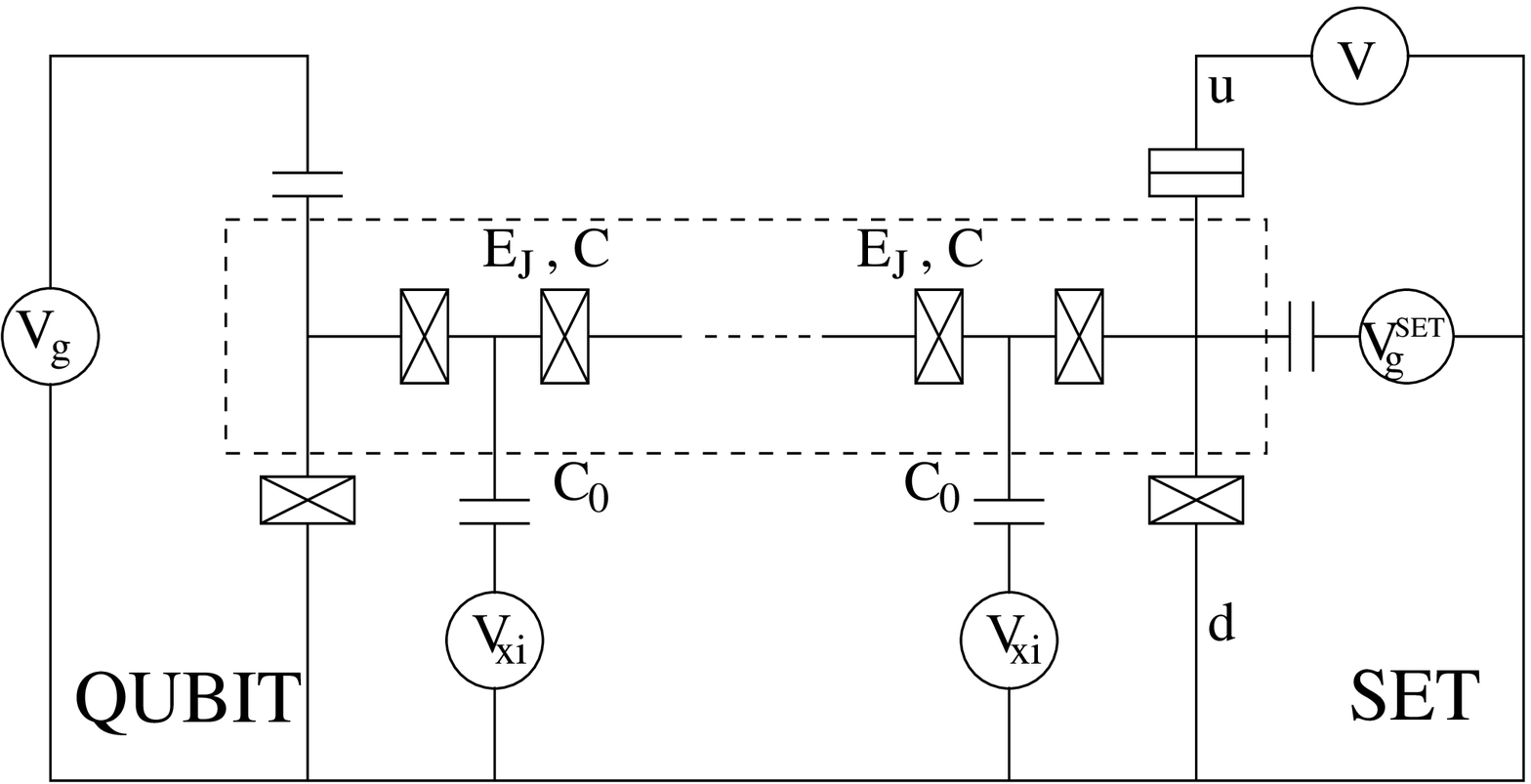}
\caption{Dashed box: one-dimensional Josephson array proposed for the 
transmission of quantum states. The crossed rectangles denote the Josephson 
junctions between the islands. The state prepared on the left-most 
island is transfered to the right-most island by the time evolution
generated by the Hamiltonian. Left part: Cooper-pair box (charge
qubit) used to prepare the state. Right part: SET transistor used as
measurement device.}
\label{fig1} 
\end{figure}

The model that we want to study is schematically illustrated in
Fig.~\ref{fig1} and described by the Hamiltonian
$
H=H_{\text{JJ}}+H_{\text{qp}}+H_{\text{coup}}
$,
where 
\begin{equation}
H_{\text{JJ}}=
          \frac{1}{2} \sum^L_{ij}  
        (Q_{i}-Q_{xi}) C_{ij}^{-1} (Q_{j}-Q_{xj}) 
        -E_{J} \sum^{L-1}_{i} 
        \cos\phi_{i,i+1}
\label{QPM}
\end{equation}
is the Hamiltonian of a one-dimensional Josephson junction 
array~\cite{faziorev} of length $L$,
and $\phi_{i,i+1}=\phi_{i}-\phi_{i+1}$.  The other terms of the
Hamiltonian describe the measurement apparatus and will be discussed
later.  The charge $Q_i$ and phase $\phi_i$ are canonically
conjugated. The first term in Eq.~(\ref{QPM}) is the charging energy,
$C_{ij}$ is the capacitance matrix; the second is due to Josephson
tunneling. An external gate voltage $V_{xi}$ gives a contribution to
the energy via the induced charges $Q_{xi} = 2eq_{xi}=V_{xi}C_{ii}$.  This
external voltage can be either applied to the ground plane or
unintentionally caused by trapped charges in the substrate (in this
case $Q_{xi}$ will be a random variable).
We assume that each island is coupled to its nearest neighbors by
junction capacitances $C$ and to the ground by a capacitance $C_0$. 
In this case, the charging interaction has 
a range given by $\sqrt{C/C_{0}}$ in units of the lattice spacing of the 
array~\cite{faziorev}. In the following we put $\hbar=k_B=1$.

In the charge regime $e^2C^{-1}_{00} \gg E_J$, the system is
approximately described by only two charge states for each island. The
chain Hamiltonian $H_{\text{JJ}}$ is equivalent to an anisotropic XXZ
spin-$1/2$ Heisenberg model~\cite{xxz1,xxz2}, the Josephson chain is
thus different from the XY and Heisenberg cases~\cite{bose03}.
It is characterized by a strong anisotropy between
the z-direction and the xy-plane. Moreover the z-coupling has a range 
which depends on the electrostatic energy and can extend over several
lattice constants. 

At $t=0$, the chain is initialized
in the state $|\bm{\psi_0} \rangle = |\psi,000...0\rangle$, where
$|0\rangle$ ($|2\rangle$) denotes the state of an island without
(with) an excess Cooper pair, and $|\psi \rangle=
\cos{(\theta/2)}|0\rangle+e^{i\phi}\sin{(\theta/2)}|2\rangle$ is the
state that has been prepared in the left-most island. This initial
state is not an eigenstate of the Hamiltonian, it will evolve as a
function of time.  In fact, as the total charge $Q=\sum_i Q_i$ is a 
conserved quantity, the dynamics is restricted to the
$L+1$-dimensional space $\mathcal{H}=\mathcal{H}_0 \oplus
\mathcal{H}_2$ of total charge zero, $\mathcal{H}_0$, and charge two,
$\mathcal{H}_2=\text{span}\{|j\rangle\}$, 
where $|j\rangle$, 
$1 \leq j \leq L$ is the state with an excess Cooper pair on the $j$-th site. 
In this basis the Hamiltonian reads 
\begin{eqnarray} 
& & H_{\text{JJ}} |j\rangle = 2 e^2 \left(
C^{-1}_{jj} -2 \sum_{i=1}^{L} C^{-1}_{ij} q_{xi} \right) |j\rangle +
\nonumber \\ & & - \frac{E_J}{2} \left( (1-\delta_{jL}) |j+1\rangle +
(1-\delta_{j1}) |j-1\rangle \right) \, .
\label{effective_h}
\end{eqnarray}

We first calculate the fidelity of transmission and the time required
for the transfer of information as a function of the coupling
constants of the Josephson chain.  The quality of the transmission is
quantified by the fidelity of the (mixed) state $\rho_L$ of the
right-most island (site $L$) to the initial state
\begin{equation}
F_L(t) =\frac{1}{4\pi}\int \langle\psi | \rho_{L}(t)|\psi \rangle d\Omega \; .
\label{fido}
\end{equation}
This definition gives the fidelity averaged over all possible
initial states on the Bloch sphere, $1/2 \le F_L \le 1$.

\begin{figure}[b]
\includegraphics[width=70mm]{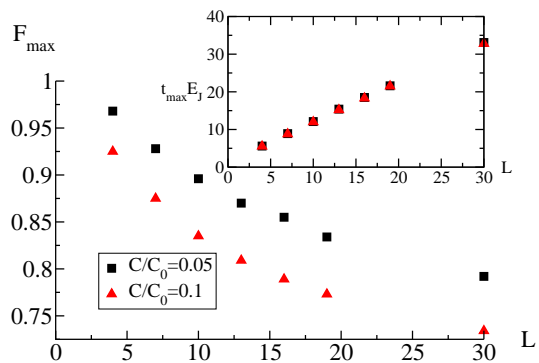}
\caption{Maximum value of the fidelity as a function of the length of
the chain for two different values of $C/C_0 \ll 1$ and
$(2e)^2/(E_JC_0)=10$. Inset: time at which the maximum is reached.}
\label{fig2}
\end{figure}

\begin{figure}[b]
\smallskip
\includegraphics[width=70mm]{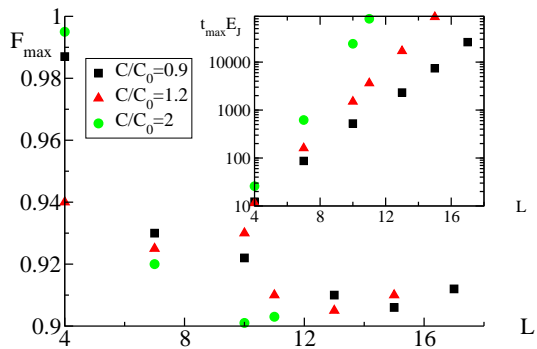}
\caption{Maximum value of the fidelity as a function of the length of
the chain for three different values of 
$C/C_0 \gtrsim 1$ and $(2e)^2/(E_JC_0)=10$.
Inset: time at which the maximum is reached.}
\label{fig3}
\end{figure}

The fidelity is a strongly oscillating function of time.  Only at
well-defined times the state is transferred faithfully through the
chain.  This does not necessarily correspond to the time in which a
Cooper pair has been transferred, since also the relative phases of
the state have to be reconstructed.  In Fig.~\ref{fig2} we show the
value of the first fidelity maximum and the time at which it is
reached as a function of the length $L$ of the array and for different
values of the ratio $C/C_0$.  For the parameters considered, the
fidelity is never smaller than 75\%.  For longer chains, or if the
condition $C_0 \gg C$ is released, the first maximum of the fidelity
is considerably reduced.  Another option is to fix a threshold for the
fidelity of transmission and seek for the first local maximum above
the threshold. The time at which these maxima occur increases
exponentially with the chain length. The value of the fidelity does
not necessarily decrease on increasing $L$, and for larger arrays a higher
fidelity can be achieved (although at larger times), see Fig.~\ref{fig3}.
The results of Figs.~\ref{fig2}, \ref{fig3} are encouraging since they
indicate that faithful state transmission using Josephson chains is
already possible with present-day technology.

\begin{figure}
\includegraphics[width=70mm]{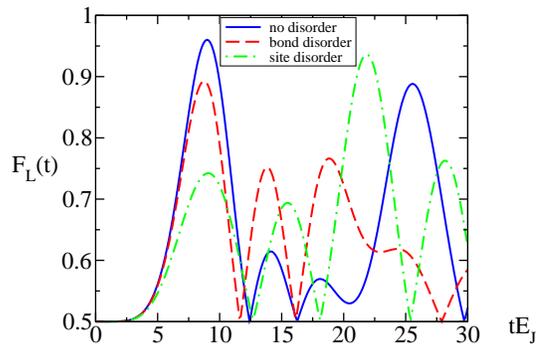}
\smallskip
\caption{Fidelity as a function of time for an array of
length $L=7$, $(2e)^2/(E_JC_0)=10$ and $C=0$, i.e. a junction 
capacitance much smaller than the ground capacitance. 
Disorder parameters: relative variance $\Delta E_J/E_J=0.1$ for bond
disorder, absolute variance $\Delta Q_x/2e=0.025$ for site disorder.}
\label{fig4}
\end{figure}

Since experimental arrays are never completely homogeneous, 
we now consider the case in which a small amount of static
disorder is present.  In general, imperfections will reduce the
fidelity. In Fig.~\ref{fig4} we show both the effect of
bond disorder (Josephson couplings distributed around an average
value) and site disorder (mimicking the effect of static background
charges and/or different capacitances).  
The effect of charge disorder appears to be more disruptive: this is 
because additional frequencies enter the dynamical evolution making 
the reconstruction of the additional wave-function more difficult.

\begin{figure}[b]
\includegraphics[width=70mm]{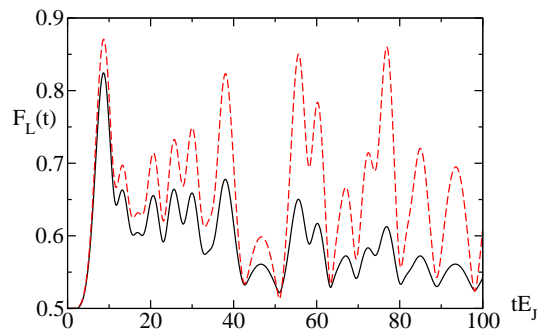}
\caption{Fidelity versus time in presence of gate voltage
fluctuations (full line). Dashed line: noiseless case.
$L=7$, $e^2/(E_J C_0)=10$, $C/C_0=0.1$, $\gamma=0.01$.}
\label{fig5}
\end{figure}

Dynamical fluctuations play a different role.  They arise from
gate-voltage fluctuations and are described by adding stochastic terms 
to the gate voltages, $q_{xi}\rightarrow q_{xi}+\xi_i(t)$ in the Hamiltonian 
in Eq.~(\ref{QPM}). Here we choose a very simple model and assume 
the $\xi_i(t)$
to be independently gaussian distributed: $\langle \xi_i(t)\rangle=0$,
$\langle \xi_i(t) \xi_j(t')\rangle=\gamma \delta_{ij}\delta(t-t')$.
Nevertheless, due to capacitive coupling between separated sites, such
stochastic factors result in correlated stochastic terms in the
effective Hamiltonian Eq.~(\ref{effective_h}),
$H_{\text{noise}}|j\rangle =H_{\text{JJ}}|j\rangle -2\Xi_j(t)|j\rangle$,
where the zero-averaging gaussian functions $\Xi_i(t)$, are uniquely
fixed by $\langle\Xi_i(t) \Xi_j(t')\rangle=\gamma
[(C^{-1})^2]_{ij}\delta(t-t')$.  Averaging out the stochastic terms
leads to the master equation for the density matrix of the chain in
the space $\mathcal{H}$, 
\begin{eqnarray} 
& &\dot{\rho}= -i[H_{\text{JJ}},\rho]\\
&-&\sum_{i,j=1}^{L}\frac{\gamma [(C^{-1})^2]_{ij}}
{8e^2} \left(Q_iQ_j \rho -2 Q_i \rho Q_j +\rho Q_iQ_j\right) \, ,
\nonumber 
\end{eqnarray}
where the operators $Q_i$ projected on the space $\mathcal{H}$ are
$Q_i=2e|i\rangle\langle i|$.  The state of the system develops into a
completely incoherent mixture in which any charged state is equally
probable, $\rho_{ii}(t \rightarrow \infty)=(1/L) \rho_{11}(t=0)$. The
averaged fidelity as defined in Eq.~(\ref{fido}) is reduced to
$F_{\infty}=1/2+1/(6L)$, corresponding to an almost unfaithful
transmission. The time dependence of the fidelity in the noisy system
is presented in Fig.~\ref{fig5} where it is compared with the fidelity
in the absence of noise. The peaks of the fidelity are not smeared out
by noise.  The dominant effect of the coupling to the environment is
the relaxation of the fidelity amplitude towards the stationary value
(independent on the initial state).  Numerically, such relaxation
takes place on a characteristic time scale $\sim 1/(L\gamma)$. Thus,
to observe high values of the fidelity it is important to have a
maximum at a short time. As a consequence, $C/C_0 \ll 1$ is
preferable, and the condition $\gamma \ll E_J/L^2 $ is required to
have a high value for the first maximum of the fidelity.

Finally we discuss how the fidelity can be measured in a practical
setup. To do this, we assume that the right-most island (site $L$) is
part of a SET transistor.  We therefore specify the effective coupling
Hamiltonian between the right-most island and the
leads~\cite{averin89},
\begin{eqnarray}
 H_{\text{qp}} &=&\sum_{b=u,d,L}
\sum_{k,\sigma} \epsilon_b(k)
\gamma^{\dag}_{k\sigma b}\gamma_{k\sigma b} \, ,\\ 
H_{\text{coup}}
&=& \sum_{b=u,d,L}\left[ e^{-i (\phi_L-\varphi_b)/2} X_b
+\textrm{h.c.} \right] + \nonumber \\ &-& \sum_{b=u,d} J_b \cos
(\phi_L -\varphi_b) -\sum_{b=u,d} V_b Q_b \, ,
\label{sset_hamiltonian}
\end{eqnarray}
where $\gamma$, $\gamma^{\dag}$ are annihilation and creation 
operators of quasiparticles in the grain $L$ and in the 
leads, $u$ and $d$ (see Fig.~\ref{fig1}). The operator 
$X_b=\sum_{k,q,\sigma}T_{qk} \gamma^{\dag}_{k\sigma b}\gamma_{q\sigma L}$ 
describes quasiparticles tunneling into the grain 
with an associated charge increasing $e^{-i(\phi_L-\varphi_b)/2}$. 
$Q_b$ is the total charge entering the chain from the up or down
reservoirs.  We assume non-vanishing quasiparticle tunneling only
across the upper junction ($eV_d=0$, $eV_u=-eV \approx -2\Delta$) and
conversely we allow coherent Cooper pairs tunneling only in the lower
junction ($J_u=0$, $J_d=J \ll E_J$).  Gate voltages are chosen so that
the SET is off resonance and we can therefore neglect the stationary
current through the SET due to the Cooper-pair quasiparticle
cycle~\cite{averin89}.

The measurement device modifies the dynamics of Cooper pairs on the
chain and requires taking into account quasiparticle excitations on
the $L$-th site of the chain.  By neglecting quasiparticle tunneling
we would have a coherent dynamics for the charges in the chain
described by the Hamiltonian $H_0 =H(T_{qk} \rightarrow 0)$.  
Tracing out the
quasiparticle degrees of freedom results, instead, in an incoherent
dynamics described by a master equation for the reduced density matrix
$\tilde{\rho}$ of charges in the chain~\cite{cohen}.  In the basis of
eigenstates of 
$H_0$, $H_0 |\alpha\rangle=E_{\alpha}|\alpha\rangle$~\cite{footnote1}, 
the master equation reads~\cite{cohen}
\begin{equation}
\dot{\tilde{\rho}}_{\alpha \beta}(t)=-i \langle\alpha| 
\left[H_0 , \tilde{\rho} \right] |\beta\rangle
-{\sum_{\mu \nu}}' R_{\alpha \beta \mu \nu} \tilde{\rho}_{\mu \nu} \, .
\label{master}
\end{equation} 
The prime indicates that the sum has to be performed over states
with energies such that $|E_\alpha -E_\beta -E_\mu +E_\nu | \ll
1/\Delta t$, $\Delta t$ being the time over which the coarse-graining
implicit in Eq.~(\ref{master}) takes place.

As we are interested in the time evolution over short times,
let us discuss in some detail our approximations.
We first assume that $J \ll E_J$, so that, in evaluating the kernel
$R$, we neglect the Josephson coupling to the leads~\cite{footnote2}.
In this case the spectrum of $H_0$ is $\{E_0,E_{1/2},E_{\bar{M}}\}$,
$\bar{M}=1,\dots,L$.  Due to the energy scale separation
$E_{\bar{M}}-E_{\bar{N}} \lesssim E_J \ll 1/\Delta t \lesssim
E_{\bar{M}}-E_{\frac{1}{2}} \sim E_{\bar{M}}-E_0 \ll eV $, the sum in
Eq.~(\ref{master}) mixes population and coherences of the density
matrix only in the subspace of $\bar{M}$ states.  In this case the
coarse-grained dynamics of Eq.~(\ref{master}) can resolve the time
scales of order $1/E_J$ that we are interested in.
In this approximation, the only non-vanishing terms of the kernel $R$ 
are found to be 
$R_{00\frac{1}{2}\frac{1}{2}} \simeq -
R_{\frac{1}{2}\frac{1}{2}\frac{1}{2}\frac{1}{2}} \simeq -\Gamma$,
$\Re e \left\{R_{\frac{1}{2}\frac{1}{2}\bar{M}\bar{N}} \right\} \simeq  
-\lambda_{\bar{M}\bar{N}}$,
$\Re e\left\{R_{\bar{M}0\bar{N}0}\right\}  \simeq  
\frac{1}{2} \lambda_{\bar{M}\bar{N}}$,
$\Re e \left\{R_{\bar{M}\bar{N}\bar{P}\bar{Q}}\right\} \simeq 
\frac{1}{2} [\delta_{\bar{M},\bar{P}} \lambda_{\bar{N}\bar{Q}} + 
\delta_{\bar{N},\bar{Q}} \lambda_{\bar{M}\bar{P}}]$,
where $\lambda_{AB}=\Gamma\langle A|L\rangle\langle L|B\rangle$.
In these expressions we approximated $\Gamma \simeq
\int_0^{\infty}\, ds \exp(i eV s) \langle X_u(s)X_u^{\dag}(0)\rangle
\simeq \int_0^{\infty}\, ds \exp(i (eV \pm
(E_{\bar{M}}-E_{\frac{1}{2}}) s) \langle X_u(s)X_u^{\dag}(0)\rangle$ as a
consequence of the separation of energy scales discussed above.  We also
neglected all other exponentially small ($\sim e^{-eV/T}$) rates.

\begin{figure}[t]
\includegraphics[width=70mm]{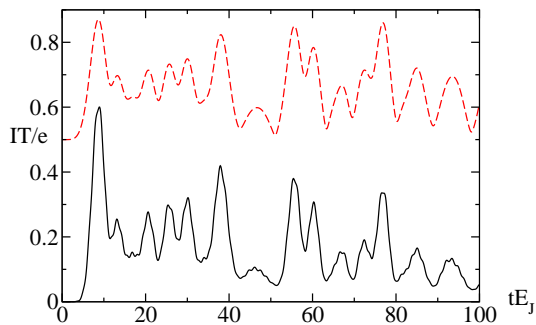}
\caption{Full line: time dependence of the current (in units of $e/T$) through 
the SET. Dashed line: fidelity of the isolated chain.
$\Gamma=0.05$, all other parameters as in Fig.~\ref{fig5}.}
\label{fig6}
\end{figure}

Finally let us address the proposed measurement protocol.  It consists
in disconnecting the right-most site from the rest of the chain at a
time $t^{\star} \ll 1/\Gamma$ and in measuring the time-integrated
current through the SET $ I=\frac{e}{T}\int_{0}^{\infty} \tilde{I}
(\tau) d\tau = \frac{e}{T} \int_{t^{\star}}^{\infty}\tilde{I}(\tau)
d\tau +\mathcal{O}(\Gamma t^{\star}) $ where $T$ is the time between
two pulses: it is the largest time scale in the system.  The
instantaneous particle current is
$\tilde{I}(t)=\Gamma(\tilde{\rho}_{LL}(t)+
\tilde{\rho}_{\frac{1}{2}\frac{1}{2}}(t))$.  The last term 
$\mathcal{O}(\Gamma t^{\star}) $ in the
current corresponds to
$\int_0^{t^{\star}}\tilde{I}(\tau) d\tau=
\int_0^{t^{\star}}(\tilde{\rho}_{LL}(t)+
\tilde{\rho}_{\frac{1}{2}\frac{1}{2}}(t)) < \Gamma t^{\star}$.  As at
time $t>t^{\star}$ the SET is disconnected from the rest of the chain
and is out of resonance, the measured current is $\sim
\frac{e}{T} (2 \tilde{\rho}_{LL}(t^{\star}) +
\tilde{\rho}_{\frac{1}{2}\frac{1}{2}}(t^{\star}))$.  This measurement
scheme does not provide a tomography for the state of the right-most
site: the measured current does not depend on the coherences of
$\rho_L(t^{\star})$, to which the fidelity is sensitive.
Nevertheless, the peaks in the current correspond exactly to the
maxima of the fidelity as shown in Fig.~\ref{fig6}.  The current decay
in time due to quasiparticle tunneling happens on a time scale $\sim
1/\Gamma$ irrespective of the length of the chain. Therefore, our
measurement scheme can be used also for long chains, the
main constraint are disorder and gate voltage fluctuations.  In this
sense the current measurement allows to check the theoretical
prediction for the fidelity of state transfer.

In conclusion, we have proposed to use a Josephson junction chain as a
solid-state quantum communication channel. 
We have analyzed the read-out process at the
end of the channel and shown that the fidelity can be directly
measured in a SET device. We have also considered the influence of static
disorder and dynamical noise. Present-day technology should allow the
realization of quantum channels of the type described here.

We acknowledge fruitful discussions with G. De Chiara,
C. Macchiavello, G.~M. Palma, and S. Montangero.  This work was
supported by the EU (IST-SQUBIT2, RTN-Nanoscale Dynamics), 
by Fondazione Silvio Tronchetti Provera, and by the Swiss NSF and the 
NCCR Nanoscience. During completion of this work we became aware of
Ref.~\onlinecite{paternostro} which discusses state transmission in
a setup using SQUID loops coupled to resonators.


\begin{thebibliography}{99}
\bibitem{footnote2} 
	We neglect the Josephson coupling
	to the leads in evaluating the kernel, but it still enters the dynamics
	of the chain through $H_0$ in the first term on the r.h.s. of
	Eq.~(\ref{master}).
\end{thebibliography}
\end{document}